# Documentation of Quality Requirements in Agile Software Development

Woubshet Behutiye[†]
M3S research unit
University of Oulu
Oulu Finland
woubshet.beutiye@oulu.fi

Pertti Seppänen
M3S research unit
University of Oulu
Oulu Finland
pertti.seppanen@oulu.fi

Pilar Rodríguez
M3S research unit
University of Oulu
Oulu Finland
pilar.rogriguez@oulu.fi

Markku Oivo
M3S research unit
University of Oulu
Oulu Finland
markku.oivo@oulu.fi

## ABSTRACT

Context: Quality requirements (QRs) have a significant role in the success of software projects. In agile software development (ASD), where working software is valued over comprehensive documentation, QRs are often under-specified or not documented. Consequently, they may be handled improperly and result in degraded software quality and increased maintenance costs. Investigating the documentation of QRs in ASD, would provide evidence on existing practices, tools and aspects considered in ASD that other practitioners might utilize to improve documentation and management of QRs in ASD. Although there are some studies examining documentation in ASD, those that specifically investigate the documentation of QRs in depth are lacking.

Method: we conducted a multiple case study by interviewing 15 practitioners of four ASD cases, to provide empirical evidence on documentation of QRs in ASD. We also run workshops with two of the cases, to identify important aspects that ASD practitioners consider when documenting QRs in requirements management repositories.

Result and conclusions: ASD companies approach documentation of QRs to fit the needs of their context. They used tools, backlogs, iterative prototypes, and artifacts such as epic, and stories to document QRs, or utilized face-face communication without documenting QRs. We observed that documentation of QRs in ASD is affected by factors such as context (e.g. product domain, and size) and the experience of practitioners. Some tools used to document QRs also enhanced customer collaboration, enabling customers report and document QRs. Aspects such as levels of abstraction, the traceability of QRs, optimal details of information of QRs and verification and validation are deemed important when documenting QRs in ASD requirements management repositories.

## 1. Introduction

Quality requirements (QRs), also referred as non-functional requirements, are prominent for the success of software projects [11]. QRs define requirements regarding quality concerns that are not covered by functional requirements (FRs) [24]. They describe the quality properties required by a system to be developed such as usability, reliability, portability and maintainability [36]. In agile software development (ASD), where late changes in requirements are welcomed to meet dynamic demands of businesses, QRs are usually underspecified or undocumented, and not considered early enough in the software development cycle as functional requirements [26]. In such cases, their mistreatment may result in project failure or loss [27].

The scientific literature of requirements engineering in ASD reveals challenges regarding the documentation of QRs. For instance, ASD user stories are insufficient to specify and document QRs [9,17,20,22], and writing acceptance criteria of QRs is challenging [25]. Approaches for documenting QRs in ASD are limited [1]. Alsaqaf et al. [2] found that ASD teams face trouble in writing Definition of Done (DoD) of QRs and that the lack of understanding of QRs created a challenge for writing the DoDs. Behutiye et al. [7], identified that documenting QRs improperly (e.g. unclear specifications, outdated and missing QRs documentation) is one of the challenges of managing QRs in ASD.

ASD advocate the continuous delivery of valuable software and minimal documentation [5], and usually favors FRs over QRs [26]. Additionally, its focus on close collaboration with customers may encourage developers to under-specify QRs [32]. Consequently, when underspecifying or not documenting QRs, ASD teams face challenges in the scalability of software [10], and the traceability of QRs [6]. Moreover, missing and insufficient documentation of QRs incur technical debt [8], forcing ASD teams experience deteriorating software quality and growing maintenance cost in the long run [10]. In certain cases, the ill treatment of QRs may even result in faulty systems that may require rework [3]. In this regard, investigating the state of the practice of documentation of QRs in ASD is beneficial as it may provide insight into how ASD companies approach documentation of QRs.

Although there are studies that investigate either QRs or documentation in ASD, those that specifically examine the documentation of QRs in ASD are few. For instance, Mendez et al. [21] examined the impact of documentation debt (i.e., technical debt that is caused by incomplete and insufficient requirements artifacts) in ASD. Robiolo et al. [28] explored the indicators of potential technical debt (*identifying QRs that have not been documented although marked as important*) and waste (*identifying QRs documented but marked as not important*), by surveying practitioners. Behutiye et al. [6], examined the challenges and practices of documentation of QRs in ASD. However, the paper did not provide in depth investigation of the documentation practices (e.g. stakeholders involved in the documentation of QRs, tools, and aspects considered in documenting QRs), or examine whether the practices for documenting QRs were similar to those of FRs. In this regard, in depth investigation of documentation of QRs and the key aspects considered in documenting QRs would help enrich the limited evidence on documentation of QRs in ASD.

Investigating the documentation of QRs in ASD, in companies that operate in different domains would provide a better insight into the QR documentation practices that ASD teams adopt to minimize the risk of inappropriate handling of QRs. Therefore, we empirically examined the documentation of QRs in ASD, to get an in-depth understanding of the existing practices and identify aspects that practitioners consider important when documenting QRs in their requirements management repositories. Hence, our research answers the following research questions:
RQ1. What are the practices for documenting QRs in ASD projects?

We investigate the documentation of QRs in ASD cases, and present their QR documentation practices, including associated tools, activities and roles responsible for documentation of QRs.
RQ2. What are the aspects that ASD practitioners consider important when documenting QRs in their requirements management repositories?

We explore the important aspects that ASD practitioners consider while documenting QRs in their requirements management repositories.

Our results show that ASD teams adopt QR documentation practices that fit their contexts. Their practices involved utilizing backlogs, iterative prototypes, and artifacts such as epics, user stories, acceptance criteria, and DoDs. Experiences of the developers, and the context (e.g. product domain and size) influenced how ASD companies document QRs. Varying stakeholders were responsible for documenting QRs (e.g. product owners, project managers, and usability designers). ASD practitioners identified that traceability of QRs, levels of abstraction, details of information and verification and validation are important aspects while documenting QRs in their requirements management repository.

The remaining sections are structured as follows: Section 2 presents related work. Section 3 presents the research approach followed in the study and Section 4 provides the answers to our research question. Section 5 discusses the findings of our work. Finally, Section 6 concludes the paper.

## 2. Related work

Regarding their capability for rapid delivery of working software and responding to changing requirements, ASD approaches have been popular and widely adopted in the software industry [29]. Nevertheless, studies reveal that ASD approaches have limitations regarding the specification and documentation of QRs [2,7,9,14,20]. For instance, the capability of ASD user stories to specify and document QRs is limited [7,9,20,22]. ASD' value of '*working software over comprehensive documentation*" is seen to encourage minimal documentation and favoring functionality over QRs, which may lead to the under-specification and neglect of QRs. This may result in customer dissatisfaction, since customers may be unaware of what the developers are doing and could not easily trust the development process [16].

QRs have elusive characteristic and are hard to define and measure [18]. These characteristics exacerbate the challenges in specifying and documenting QRs in ASD. Alsaqaf et al. [2], investigating the challenges of QRs in large scale distributed ASD, identified that ASD teams experience difficulties in specifying DoDs of QRs, writing test specifications of QRs, and in precisely specifying QRs. According to their findings, unclear conceptual understanding of QRs may lead to ambiguously specifying QRs in user stories and DoDs. Additionally, they found that minimal documentation might result in missing the rationales behind QR tradeoffs and architecture decisions taken earlier. In a recent systematic mapping study of management of QRs in ASD and rapid software development, we [7] identified that QR documentation challenges may arise from unclear and missing QR documentations, and the difficulty in ensuring end-to-end documentation of QRs.

Although there are studies that investigate documentation in ASD [6,15,21,28,33–35], those that specifically examine the documentation of QRs are few. Additionally, detailed investigation of practices of documentation of QRs in ASD is missing. For instance, Hoda et al. [15], investigated documentation practices in ASD. They found that ASD teams apply electronic backups of physical paper artifacts, document change decisions made by customers, business terminologies and functional specifications and customers' feedback. They also revealed that ASD teams that relied on paper artifacts (user stories written in cards, and post it notes), experienced challenges (e.g. losing data and time). However, their study did not address the documentation of QRs. Stettina et al. [34] examined the impact of documentation formalism on developers' documentation practice in ASD. They found out that documentation was seen as intrusive task and was often assigned to less qualified team members. They also found that iterative documentation practices and following formal document templates enabled capturing detailed development knowledge.

On the other hand, some studies investigating documentation of QRs in ASD, focused on documentation debt. Soares et al. [33] examined the difficulties of user stories in ASD and analyzed whether the difficulties were related to documentation debt. They reported eight difficulties, among which the lack of information and identification of QRs, were causes of documentation debt. Mendez et al. [22] investigated the impact of documentation debt in ASD. They identified that the lack of QRs identification and the lack of information were related to high proportions of documentation debt. Robiolo et al. [28], explored the indicators of technical debt and waste resulting from QR documentation based on survey findings.

Behutiye et al. [6] identified that ASD companies applied varying practices to document QRs. For instance, ASD teams in small and medium sized companies favored face-face communication and kept the need for documentation of QRs minimal, while in large sized companies, ASD teams utilized multiple and complex backlog structure to document QRs. However, the study did not provide detailed analysis of the practices regarding documentation of QRs, or the key aspects practitioners consider during QR specification. Therefore, regarding the significance of QRs and the limited evidence on how QRs are documented in ASD, we aim to explore the existing practices of ASD companies in documenting QRs.

## 3. Research approach

Our study focuses on examining the state of the practice of documentation of QRs in the context of ASD. We adopted the guidelines for conducting and reporting case studies by Runeson and Höst [30], to investigate the QR documentation practices in ASD through multiple case study of four cases. Case study is best suited for investigating a specific phenomenon in its context [30]. For the purpose of the study, we developed and applied a case study design protocol, which has been reviewed by experienced researchers prior to starting the study. The protocol formulated the study objectives, research questions, data collection methods and selection strategies.

In the following sections, we describe the steps followed in designing and executing the study. Section 3.1 presents the case and participant selection. Section 3.2 provides the data gathering procedures. Section 3.3 presents the data analysis process.

### 3.1. Case and participant selection

We selected four cases that employ ASD, in order to gather information on QR documentation practices in ASD. These cases varied in terms of their sizes, products and geographical location, providing us an opportunity for examining the state of the practice of documentation of QRs in a wider context. Table 1, presents the summary of the cases in our study.

**Table 1. Summary of the cases**

| Case | Software development approach | Product domain | Company size in terms of employees |
|---|---|---|---|
| A | ASD | Modelling tool | Over 900 |
| B | Scrum based ASD | Telecommunication and embedded systems | Over 600 |
| C | Large scale distributed ASD | Telecommunication | Over 100,000 |
| D | Scrum based ASD | Web application | Less than 100 |

We used the key informant technique [19], in order to recruit the participants in our study. The key informant technique provides a means for collecting quality data by using experts as sources of information on a topic [19]. For this purpose, we contacted the champions of the four cases, and informed them about the objective of our study and potential roles that might be participants of our study. While proposing the potential roles, we consulted ASD literature and as well as baseline stakeholders in requirements engineering suggested by Sharp et al. [31]. We proposed practitioners (those who are involved in the development process such as developers, testers, quality assurance engineers), and decision makers within the organization (e.g. project managers, product owners, analysts, and release engineers). Our rationale in proposing the roles was to get rich and relevant information on the topic, as they are involved and affected by the requirements engineering process. Following this, the champions selected and helped us in recruiting subjects with relevant skills and knowledge to be participants in our study.

### 3.2. Data gathering

We used semi-structured interviews to collect data for answering RQ1. Furthermore, we conducted a workshop with cases B &C to get an in depth understanding of the important aspects they consider while documenting QRs in requirements management repositories, to answer RQ2.

*3.2.1 Semi structured interviews*. We collected data regarding documentation of QRs through semi-structured interviews. There were 15 interviewees, from the four cases. We asked the interviewees questions regarding the QR documentation practices employed in their company, also including tools, artifacts and roles involved in documenting QRs. The interviews were audio recorded and later on transcribed for data analysis purpose. Table 2 summarizes the participants' role and experience.

**Table 2. Interview participants**

| ID | Interviewee role | Case | Experience (years) | ASD experience (years) |
|----|------------------|------|---------------------|------------------------|
| 1  | Project manager  | A    | 20                  | 10                     |
| 2  | Software developer and architect | A | 11 | 11 |
| 3  | Executive manager | A   | 30                  | 13                     |
| 4  | Production test lead | B | 25                | 5                      |
| 5  | Technical lead   | B    | 15                  | 15                     |
| 6  | Project manager  | B    | 19                  | 12                     |
| 7  | Process coach    | B    | 15                  | 6                      |
| 8  | Line manager     | C    | 3                   | 3                      |
| 9  | Quality lead     | C    | 24                  | 12                     |
| 10 | Transformation expert | C | 1.5               | < 1year                |
| 11 | Quality manager  | C    | 25                  | 10                     |
| 12 | Software engineer | C   | 6                   | 6                      |
| 13 | Quality manager  | C    | 18                  | 10                     |
| 14 | Software engineer | C   | 16                  | 6                      |
| 15 | Product owner and chief software architect | D | 10 | 5.5 |

*3.2.2 Workshop.* We conducted follow-up workshops with cases B and C, in order to get in depth understanding of the significant aspects ASD practitioners consider while documenting QRs in their requirements management repositories and as well complement our initial findings on QR documentation practices. This was done after the semi-structured interviews, which provided insights into the QR documentation practices of the cases. Table 3, summarizes number of the participants and duration of the workshop sessions.

**Table 3. Summary of workshop sessions**

| Case | Participants | Participant roles | Duration of workshop |
|------|--------------|-------------------|----------------------|
| B    | 2            | Tech lead, Senior engineer | 90 minutes |
| C    | 3            | Project manager, Software engineer, Transformation expert | 85 minutes |

The workshops were conducted face-to-face with the participants, who are agile practitioners in the companies, as follows:

1. First, we presented the objective of the workshop to the participants. The objective was to determine the significant aspects considered by ASD practitioners when documenting QRs in their requirements management tools and as well as corroborating our understanding of their QR documentation practices.
2. As Jira was a requirement management repository used in both cases, a generic Jira template for documenting QRs, was presented to the participants to initiate discussions on how QRs are documented at different levels of abstraction. The generic Jira template was prepared based on a consultation with a process coach in case B and the first author, before conducting the workshop.
3. Participants were asked to discuss and reflect up on important aspects they identify when documenting QRs in their requirement management tool. These were the important aspects that they considered mandatory for optimal documentation of QRs at the respective level of abstraction (e.g. Epic, story, task).

We recorded audio of the workshop sessions and transcribed the recordings for the purpose of analysis.

### 3.3. Data analysis

As we had collected data in two steps for the two RQs, we applied the data analysis separately. We explain the data analysis steps for analyzing the data from the semi-structured interviews and workshops as follows.

*3.3.1 Data analysis of semi-structured interview data.* In order to analyze the data, we first coded the transcribed documents in NVivo, a qualitative data analysis tool. We applied deductive and inductive coding approaches [12] and labeled the transcriptions. Then, the related labels with similar concepts were categorized together to identify themes. Thus, we applied thematic analysis to determine the QR documentation practices in the cases [12].

*3.3.2 Data analysis of workshop data.* To master the wide-ranging research data collected during the workshop, we opted for conducting a qualitative analysis using in parallel the thematic synthesis and narrative synthesis methods presented by Cruzes et al. in [13]. The narrative synthesis method highlighted the case specific variations, while the thematic synthesis helped us to identify commonalities and draw conclusions.

The narrative synthesis was started by reading through the transcriptions and labeling the sections containing data relevant to answering the research question, with codes. This resulted in a set of case-specific findings. The narrative sections were copied to Excel spreadsheets along with the codes without modifying the content of the narratives. These findings were further cross-analyzed in order to identify common themes, providing the basis to answer RQ2. The analysis was first conducted by the second author, and the results were reviewed and refined by the first author.

### 4. Result

### 4.1. Practices for documenting QRs in ASD

**Case A** applies ASD and is as well experienced in model-driven development. Practices for documenting QRs in the case varied depending on the context and the need. For instance, while planning features during release planning meetings, the executive manager together with the product manager and project managers discuss and document QRs such as performance, and user experience in word documents. A response from the executive manager shows the flexible approach regarding the documentation of QRs. He stated that when it comes to documentation, *"There is no single practice. The practice may vary according to, the target and the context. It could be whiteboard meetings or other practice. On some evolutions, the technology and the solution may be unclear and, what we should expect from it may be unclear. So we may have an iterative prototype process, which helps us, both to discover*

*what will be the best architecture to, adopt and also what we can expect, as performance, as quality aspects, as usability, before we can write any reasonable set of requirements. Sometimes we are, accurate enough to do a specification document with a requirement list, in a precise way, in order to implement properly the solution.*" The case also applies models (e.g. UML models), which serve as a means for communicating QRs. Moreover, team members' interaction and minimal documentation are also a focus in the case. As a result, while discussing implementation of features, the teams rely on face-to-face and white board meetings and document decisions regarding QRs in word documents. While there are not templates for documenting QRs, developers can document QRs as user stories in word documents or in other formats, based on their decision during the white board meetings.

Developers, project managers, product managers, development managers and sales team document QRs during the software development lifecycle. Project managers and developers document QRs related to maintenance issues in Redmine, which is an issue tracker tool used internally in the case. Similarly, customers report and document QRs and other quality issues in Mantis tool.

**Case B** follows Scrum based ASD approach. It uses QR documentation practices such as using an issue tracker tool to document QRs, documenting QRs in ASD artifacts, and applying guidelines to help with documentation of QRs. The case applies a requirements management guideline that provides detailed information on QRs and recommendations on how to document them in the backlogs. The guideline lists types of QR (e.g. security, usability, testability) and provides examples on how to document them. Depending on the type of the QR, the case applies additional guidelines when specifying and documenting QRs. For instance, specifying and documenting security requirements requires considering security standards and certifications.

The case uses Jira, an issue tracker tool, to document both QR and FRs. It also uses an agile playbook that describes recommended practices for developing software, including practices in using Jira to document both QRs and FRs. In general, it uses artifacts such as epics, stories, and tasks, to represent the levels of abstraction of QRs and document them in backlogs, in Jira. QRs documented as tasks are linked to stories and the stories are linked to epics. Additionally, it documents QRs resulting from legacy errors with error labels in issue trackers. Jira templates for documenting both QRs and FRs are similar. Documenting the QRs in these templates covers aspects such as describing the QR, the verification method for the QR and a DoD, which defines an exit criterion for the QR. Specifying the DoD for the QRs at epic level may comprise many exit criteria that apply to multiple stories. For instance, it may include stating that stability testing to be done, and meeting a specific percentage of test coverage. On the other hand, specifying QRs in DoD at story and task levels may follow a recommended structure in the form of "*Given/when/then*" to fulfil the needs of the specific user story or task. For instance, a DoD for reliability QR can be specified as follows at story level, "*Given* that the system is in a non-functioning state*, when* applying the fixes*, then* the system should reach a normal steady state*".*

Product managers, product owners and developers are involved in documenting QRs during the software development process and use Jira, word documents and prototypes to document QRs. Additionally, being a telecommunication and embedded systems development company, the case applies separate organizations that are responsible for documenting specific QR types. This was explained by one of the interviewees: "*We have some kind of categorization of those, and the organization is somewhat also split, based on the focuses we have. The security domain who work for multiple projects, they are in charge of the security architecture of multiple products and produce the relevant documentation for those. For the performance, it goes maybe to more on the test automation side but even there we have the specific people who are, just checking the performance*".

**Case C** applies large scale, distributed ASD. It has varying practices to document QRs within the software development process depending on the organizational level and the type of QR (e.g. security, performance). In general, there are multiple backlogs to document both QRs and FRs. The case documented QRs in requirement management tools like Focal Point, DOORS, Accept360, and in pronto, a bug tracker tool and within backlogs in Jira. Some development teams at lower level also utilized offline post it notes as requirements backlogs to document both QRs and FRs. Generally, the backlogs at the case were structured in such a way that there were multiple lower level backlogs which were documented in multiple tools, being independent of upper level backlogs. However, the case is in transition towards a backlog structure where lower level backlogs are inherited from one main upper level backlog with Jira. In addition, the case follows additional standards while specifying and documenting QRs like security and performance.

In Jira, the case applies levels of abstraction: features, system items, entity items, competence area items, epics, tasks, and sub-tasks while documenting QRs. QRs are also documented as DoDs of tasks describing the exit criteria. However, while DoDs are applicable in cases where the task is mainly dependent on software, they are not used in cases where implementation of the task is dependent on hardware requirements.

During the development process, the case also utilizes a special backlog, "improvement backlog", where improvement ideas during the development process are documented and tracked. QRs are also documented in the improvement backlogs, as shown in the response from the quality lead, "*We have improvements of all kind in this improvement backlog. There we have ideas from anyone. Everybody can put an improvement idea in the backlog. It is everything else but not the feature. We have not limited anything so it can be something like, I don't like this color, I like the red one or, it can be something like we need to improve our unit testing code coverage from 50 percent to 95 percent, or whatever*". However, the improvement backlogs are separated from product backlogs' items. As a result, implementation of QRs, documented as improvement backlog items, depends on how teams handling the improvement backlog are pushing the improvement actions to product backlogs.

Depending on the level of the organization, roles like managers, product owners, scrum teams and dedicated teams specify and document QRs. For instance, dedicated teams are responsible for the system level specification of FRs and QRs, whereas product owners document and handle QRs as sprint backlog items. Additionally, customers also report and document QRs, FRs and feature requests in Focal point.

**Case D** applies Scrum based ASD approach. In general, the case documents QRs together with FRs. It uses word documents, mockups and software development repositories to document QRs. For instance, it uses Sketch tool to specify and handle usability and user experience aspects. It also documents QRs as DoDs or acceptance criteria of FRs, which are written as user stories.

Moreover, the product owner and scrum teams work closely which facilitates clear communication on both QR and FRs. When experienced developers are involved in these interactions during development, the need for strict specifications of QRs may not be necessary as they are aware of the QRs, and specify them. The product owner reflects this in a response, "*for example if you work in a project for half a year or for three or four months, there's always an initial phase that you need to describe for example how the tool should look like. In addition, you give the developers the UX design and so on, using Sketch and other tools like that, and they specify for example the paddings and the margins of the specific elements of your user interface. But afterwards, for further user interface, for further, I don't know, models and views, you don't have to specify each time the paddings and margins and so on.*" The product owner is mainly responsible for handling the specification and documentation of QRs. However, the input from sales team, developers and analytics team support his decisions when specifying and documenting requirements (both FRs and QRs). For instance, the analytic teams provide a requirements specification document, which specifies the FRs and how these FRs should be working, and the higher level QRs. The Product owner further analyzes and uses the document to specify QRs. Table 4 summarizes the documentation of QRs in the cases.

**Table 4. Summary of documentation of QRs in the cases**

| Case | Tools and artifacts used to document QRs | Practice overview | Roles documenting QRs |
|---|---|---|---|
| A | Mantis, Redmine, Word document, iterative prototypes, models, whiteboards | Minimal documentation, relied on face-face communication | Product manager, project managers, developers, sales team, customers |
| B | Jira, epics, stories and tasks, DoDs, acceptance criteria, verification methods | Agile play book, guidelines to document QRs, separate organizations handling specific QRs | product owner, product managers, developers |
| C | Jira, Focal point, DOORS, Accept 360 customer feature, change request, internal system feature, DoDs | Distinct practices based on organizational level and type of QR (e.g. separate organizations documenting QRs, using post it notes at lower level) | Managers, product owners, scrum teams and dedicated teams and customers |
| D | Word document, mockups, Sketch, DoDs of FR | QRs as DoDs of FRs, No need to document QRs if there are experienced developers | Product owner |

## 4.2. Important aspects when documenting QRs in requirements management repositories in ASD

The findings from the workshops with cases B and C, reveal four important aspects that agile practitioners consider while documenting QRs in their requirements management repository. These are the levels of abstraction, the traceability of QRs, optimal details of information of the QR and verification and validation aspects. We present these aspects as follows.

*4.2.1 Levels of abstraction.* Employing levels of abstraction while documenting QRs is an important aspect considered in both cases, B and C. The levels of abstraction refer to the granularity levels used to represent the requirement. Case B applies levels of abstraction while documenting requirements (both QRs and FRs). The case documents QRs in multiple backlogs. In general, at higher levels, QRs are documented in the main requirements backlog. This is further refined in the product requirements, as product requirement epics. Within the product backlog, QRs are represented at three levels of abstractions as epics, stories and sub-tasks. An epic represents higher levels of requirements and it is a grouping of several stories, which in turn are split into multiple tasks. For instance, a participant reflects up on how QRs can be split from epics to stories as follows: "*Let's say we have code quality or code style epic and you have a QR that points out that this area in the code should be documented better. Then you might have a testing epic and then things like increased unit test coverage goes below that, or you might use to have a QR, very big epic that all the QRs fit in under somewhere, in that you have probably some way of organizing them between the story and the epic*". Similarly, Case C applies levels of abstraction for documenting QRs. However, due to the large size, complexity and variety of backlogs and as well as the large-scale distributed nature of company, the levels of abstraction for documenting QRs and FRs differ depending on the selected backlog. For instance, Feature items (consisting of both QRs and FR items) are documented in the feature backlog, in Jira. Then, features from the feature backlog are refined and specified into system item, entity item and competence area items in decreasing order of level of abstraction, within the product backlog, where items from all products are stored in a single project in Jira. On the other hand in competence area backlogs, the competence area items (QRs and FRs) are specified following Epics, story (task) and sub-task hierarchy. Both cases apply the levels of abstraction while documenting QRs and FRs.

*4.2.2 Traceability of QRs.* Ensuring the traceability of QRs is an important aspect while documenting QRs. Since QRs are documented at different levels of abstraction, keeping the link between these levels is important to ensure the traceability of QRs. Case B documents the links between levels of abstraction of QRs (e.g. Epics are linked to product requirement themes, stories linked to epics, and tasks are linked to stories) to keep traceability.

One of the participants highlighted the importance of traceability among the levels of abstraction as follows: "*For the story I have always created tasks. Because, if I am not creating tasks for a story type and I am creating just technical tasks then I need to manually add the link, but either way I will add the link. Because the hierarchy needs to be there for traceability*."

In Case C the traceability of QRs is established by linking requirements at varying levels of hierarchies among distinct backlogs and within each of the backlogs. For instance, traceability among epics, stories and subtasks, within a competence area backlog is ensured by linking subtasks to stories, and linking the stories to epics. Additionally, an epic in one backlog can be linked to epics in other projects when they are related, and the access and visibility to the backlogs of other projects is possible. One of the participants highlighted the importance of traceability as follows: *"And also if you think that, in the project, we have discussed about tracing back, meaning that if we are able to trace back from the certain backlogs to requirement documentation to features to whatever. So basically if in the development phase, we violate something, some quality requirement we already have, we should be able to trace back what requirement we are violating with certain choices. So for that reason we would also need the ID."*

*4.2.3 Optimal details of information on the QR.* Another important aspect that received attention within the two cases is the level of detail of information conveyed in the artifacts (e.g. Epics, stories and tasks). The cases adopt Jira templates, which are tailored for specifying QRs at different levels of abstraction.

In Case B, the Jira templates consist of mandatory fields that developers and product owners must fill while specifying QRs, to ensure optimal documentation. These fields may vary depending on the level of abstraction. For instance, specifying QRs at Epic level requires specifying the method of verification (e.g. customer review, design review, test case) besides other properties (e.g. description, priority, DoDs and linked issues). However, this is not necessary while specifying QRs at story level. A participant reflects up on the variations of the fields as follows: "*Yeah because the product requirement level is about, it is about the product owners, like Mr. X, was saying that we need to see a list, have we done everything so that is why there is a field to check that, in what phase did we do the check. But, for this story level which is already about the team, about the implementation, then we trust our DevOps process, we trust the process, in the process we use Gerrit, Jenkins, code reviews, the quality is already built in. We don't need to say that, verification method is, during coding with your partner, no, it is, whatever method, but it needs to happen during development.*" Table 5 presents an example of maintainability QR specified at story level. As shown in the table, the case, for instance, enforces specifying summary of the story, its description, related components, DoD, linked epics and priority.

While specifying stories, the case applies a user story template, which focuses on communicating the relevant stakeholder, the required task, and the expected outcome. For instance, a participant provides a user story about the reliability of asset tracking (i.e., reliability of different software components that handle data) as follows: "*As a user I want to avoid accidentally sending emergency messages*". Moreover, while specifying DoDs, practitioners may use the 'Given/when/then' template to convey the exit criteria required for QRs to be marked as done. However, although the template is recommended practice it is not strictly followed.

**Table 5. QR specification at story level**

| Field | Description |
|---|---|
| Project | XYZ |
| Issue type | Story |
| Summary | Complexity of files should be below 20% |
| Component | SW component Y |
| Reporter | Mr. Z |
| Priority | Minor, Medium, **Major** |
| Description | The Q analysis found that the percentage of complex files was above 34% |
| Definition of Done | The amount of complex files should be below 20%. |
| Epic link | The linked epic e.g. epic_code_quality |
| Due date | 28.12.2019 |

In Case C items in Jira are configured depending on the needs of the specific backlogs. While the general level of abstraction for QRs as epic, story and subtask is followed, QRs can also be documented as, "*features, change requests, internal features, system items, system technical analyses, entity items, entity technical analyses, competence area items and epics*", as noted by one of the participants. Jira templates in the case, consist of distinct fields that need to be filled in while documenting QRs as epics, tasks and sub-tasks. For instance, fields at epic level comprise the related project, epic name, summary, and description, specifying priorities (business level, product management level), assignee and reporters, DoDs, and QA fields (whether API review, code review and CI tests have been done/not done). In addition, in some teams it is possible to add new fields that complement the default ones. However, one participant pointed out the importance of ensuring that the new fields will not duplicate properties of already existing fields, as follows: "*of course if you have a great number of special fields, because Jira anyway has to support like default fields and then if you duplicate all the fields, you can imagine that you can get some kind of performance issues with the tool. Especially if the number of issues are also, a lot. Let's say so. You are measuring those in hundreds of thousands of issues, recorded in Jira I would say.*"

*4.2.4 Verification and validation aspects.* Another important aspect in specifying and documenting QRs in JIRA backlogs within the cases, is verification and validation. We observed that both cases incorporate fields that help ensure including aspects of verification and validation for the QRs. The details and options of verification fields varied in each of the two cases, as their processes vary. Moreover, the methods for verification and validation differed along the levels of abstraction of the QRs.

In Case B, specifying QRs at Epic level requires filling in the verification method for fulfilling the epic requirement (e.g. design review, customer review) and specifying the DoD which define the exit criteria of the epic. On the other hand, at story and task levels, the DoD is used for validating the related QRs.

Case C adopts distinct fields that support verification and validation of the QRs at different levels of abstraction. At Epic level, methods for verifying the QRs (e.g. API review, code review, and continuous integration tests) need to be filled in and serve as verification method. Similarly, DoDs provide a means of validating QRs at epic level. At story and subtask levels, QRs are documented in DoDs, and similar to case B, details regarding the verification method do not have to be specified at these levels.

## 5. Discussion

### 5.1. QR documentation practices in ASD

We observed that the cases apply practices that they deem suitable for their context while documenting QRs. For instance, the cases operating in the telecommunications domain applied separate internal organizations that are responsible to document and handle specific QRs (e.g. security). Such structure may affect the agility of the process, as it enforces additional documentation needs. However, it is essential in order to meet the regulatory requirements needs in the domain. On the other hand, in the cases operating in web and modeling application domains, the need for the separate organization was less. These cases did not have separate organizations handling specific QRs. However, they had small ASD teams working closely, and used face-to-face communication and utilized whiteboards and flipcharts instead of formal documentation of QRs. This was suitable in the contexts as there were not strict regulatory needs.

In our study, two cases documented QRs in Jira. Additionally, we found tools such as Accept 360, DOORS, Redmine and Mantis, for documenting QRs. These tools served the need for documenting and managing QRs from all available sources and stakeholders. For instance, customers reporting QR issues used Mantis to document QRs. Developers used Redmine to document and track issues, including QRs. The distinction of tools used for documenting QRs in the cases, may arise from their specific needs, e.g. context, the scale and size of their products. For instance, Focal point, Accept 360, Pronto, and DOORS were used in the large scale distributed software company, which is in the telecommunications domain. We also observed that the tools in the cases support ASD. For instance, customer collaboration is enhanced by using Focal point and Mantis tools, helping customers to report and document QRs and other issues. ASD teams can adopt these tools to support documentation and management of QRs. They can also learn the importance of covering QR documentation needs of multiple stakeholders (e.g. developers, and customers). Regarding the use of 'improvement backlog' to document QRs, we noticed that the likelihood of implementing QRs documented in such backlogs, which are separated from the main product backlog, is dependent on how the ASD teams were pushing for the improvement actions. Therefore, when adopting similar practices, it is important to consider the implementation actions.

The cases applied artifacts such as epics, stories, user stories, tasks and DoDs for documenting QRs. We found practices such as 'Given/why/then' structure for writing the DoD of QRs. This practice can be beneficial to ASD practitioners as the difficulty of writing DoDs of QRs is a challenge in ASD [2]. We also noticed that writing DoD for QRs might not be applicable in cases where the tasks are dependent on hardware requirements. Practitioners may also apply iterative prototypes to discover and document evolving QRs in ASD, as indicated in one of the cases. This practice aligns with ASD' nature of responding to changing requirement needs and minimal upfront planning of requirements.

One of the cases in our study applied guidelines to support the documentation and management of QRs. Such guidelines provide a means for clarifying QRs, their significance and the way to document and manage them. We believe that companies may benefit from such guidelines, considering challenges in managing QRs in ASD reported in the literature (e.g. the lack of awareness of QRs, difficulty in specifying and documenting QRs). Inexperienced and new ASD developers may also find such guidelines for documenting and managing QRs, beneficial.

We observed that the experience of developers might affect the documentation of QRs. For instance, a product owner in case D reported that QRs may not need to be specified while working with small sized, and collaboratively working team of experienced developers. Experienced developers were assumed to know the QR needs (e.g. usability and security). On the other hand, when there are new and inexperienced developers within the team. QRs had to be specified and documented.

We identified that the stakeholders (e.g. developers, product owners, software architects, project managers, product managers, customers) may document QRs in various stages of the software development lifecycle. However, in the case of a smaller company only the product owner was responsible for documenting QRs. In this case, other stakeholders (e.g. sales team and developers) only provided inputs on QR aspects but were not documenting QRs.

In our study, two cases approach the documentation of QRs in a similar fashion as FRs, and they explicitly document the QR details. This varied from the practices reported in the other two cases, where QRs were either not documented, or documented less compared to FRs. The context (e.g. team composition, team size, product size, and product domain) may have influenced the way QRs are documented and treated in the cases.

### 5.2. Important aspects when documenting QRs in requirements management repositories in ASD

Applying the levels of abstraction while documenting requirements is important in improving the communication and understanding of the requirement problem [23]. In our study, it was one key aspect that practitioners considered while documenting QRs. We found that the levels of abstraction were applied and that they differed depending on the details of information of QR needed at the specific stage of the software development cycle and as well according to the needs of the stakeholders in the corresponding software development stages.

The lack of traceability of QRs is a challenge in ASD [4]. In our study, keeping the traceability of QRs, either among the levels of abstraction or dependent backlogs was one key aspect in documenting them, according to the practitioners. The finding is interesting, as it may also help address the challenge arising from the lack of traceability of QRs in ASD.

Another key aspect when documenting QRs in ASD is optimal details of information on the QR. We observed that the detail of the information varies depending on the level of abstraction of the QR. The QR detail was also corresponding to the needs of stakeholders involved in documenting QRs. We find optimal details of information regarding QRs important, as it may also help address documentation debt resulting from the lack of information of QRs in ASD [33].

The verification and validation aspect was also identified important in documenting QRs. This is interesting, as verification of QRs is a challenge in ASD [7]. We believe that considering such aspects while documenting QRs may help in addressing the challenge of verifying QRs in ASD.

We observe that in both cases JIRA have been used to document both FRs and QRs. The finding reveal applicability of the tool to document and manage QRs in ASD.

### 5.3. Threats to validity

Construct validity: we applied operational measures to ensure common understanding on concepts included in our study. During the interviews, we clarified concepts and our questions to the participants to minimize threats from misunderstanding. For instance, when referring to QRs we clarified to the practitioners that we were referring to non-functional requirements, and provided examples such as usability, maintainability, and security.

Internal validity: In order to mitigate threats from internal validity, we applied triangulation through multiple data sources (e.g. workshop and additional documents from the cases) to corroborate and complement our findings.

External validity: although our findings reflect practices regarding the documentation of QR in ASD, it is difficult to generalize to other contexts. However, we believe that the findings can be extended to similar contexts applying ASD. Regarding the important aspects in documenting QRs in requirements management repositories, our findings relied on discussion initiated by using JIRA templates. Selecting another requirements management tool and template may have had different outcome.

Conclusion validity: we collected data systematically using interview scripts and audio recordings in workshops. Additionally, a second researcher reviewed and refined the data analysis results to minimize threats from subjective evaluation.

## 6. Conclusion

The paper explored and presented empirical findings on QR documentation practices in ASD companies. We identified that documentation of QRs in ASD differed depending on the chosen context (e.g. domain, team composition, size of product). We observed that cases in small sized companies applied whiteboards and flipcharts, or documented QRs as part of DoD of user stories of FRs and relied on face-face communications. In larger companies, QRs and FRs were documented in a similar way in the requirements management repository. The cases in our investigation applied artifacts (e.g. epics, stories and tasks, prototypes), and tools (e.g. Jira, DOORS, Focal point) to document QRs. Different roles were also responsible for specifying and documenting QRs. Additionally, the experience of developers influenced the documentation of QRs in ASD.

ASD practitioners valued the traceability of QRs, levels of abstractions, optimal detail of information on QRs, and verification and validation aspects, when documenting QRs. The study supports the findings from scientific literature that reveal the importance of QRs and the need for optimal documentation of QRs. In the future, we would like to expand our work by investigating in detail other factors affecting documentation of QRs in ASD and provide recommendations for optimal documentation of QRs in ASD.

## ACKNOWLEDGMENTS

This work is partially funded by the Q-Rapids project, European Union's Horizon 2020 research and innovation funded program under grant agreement N° 732253. We would also like to acknowledge champions in the case companies for facilitating the studies.